\documentclass{emulateapj}

\usepackage[latin1]{inputenc}  
\usepackage[english]{babel}    

\slugcomment{Recieved 2006 December 14; accepted 2007 January 18}
\shorttitle{Far-Infrared detection of H$_2$D$^+$}
\shortauthors{Cernicharo, Polehampton \& Goicoechea}

\begin{document}

\title{Far-Infrared detection of H$_2$D$^+$ toward Sgr B2\altaffilmark{1}
}

\author{J. Cernicharo}

\affil{
Departamento de Astrof\'{\i}sica Molecular e Infrarroja, Instituto
de Estructura de la Materia, CSIC, C/ Serrano 121, 28006, Madrid.
Spain}
\and
\author{E. Polehampton\altaffilmark{2}}

\affil{Space Science \& Technology Department, Rutherford Appleton
Laboratory, Chilton, Didcot, Oxfordshire, OX11 0QX, UK}

\and
\author{J.R. Goicoechea}

\affil{LERMA-LRA, UMR 811, CNRS, Observatoire de Paris et \'Ecole Normale Sup\'erieure,
24 rue Lhomond, 75231 Paris Cedex 05, France}

\altaffiltext{1}{Based on observations with ISO,
an ESA project with instruments funded by ESA Member States 
(especially the PI countries: France, Germany, the Netherlands 
and the United Kingdom) and with participation of ISAS and NASA.} 

\altaffiltext{2}{Department of Physics, University of Lethbridge, 4401 University Drive, Lethbridge, Alberta, T1J 1B1, Canada}

\begin{abstract}
We report on the first far--IR detection of H$_2$D$^+$,
using the  \textit{Infrared Space Observatory},
in the line of sight toward Sgr~B2  in the galactic center.
The transition at $\lambda$=126.853 $\mu$m connecting
the ground level of o-H$_2$D$^+$, 1$_{1,1}$, with
the the 2$_{1,2}$ level at 113 K, is observed in absorption against the
continuum emission of the cold dust of the source. The line is broad,
with a total absorption covering 350~km~s$^{-1}$, i.e., similar to that
observed in the fundamental transitions of H$_2$O, OH and
CH at $\sim$179, 119 and 149 $\mu$m respectively. For the
physical conditions of the different absorbing clouds
the  H$_2$D$^+$ column density ranges from 2 to 5$\times$10$^{13}$ cm$^{-2}$, i.e.,
near an order of magnitude below
the upper limits obtained from ground based submillimeter telescopes. 
The derived H$_2$D$^+$ abundance is of a few 10$^{-10}$,
which agrees
with chemical models predictions for a gas at a kinetic temperature of
$\simeq$20~K.
\end{abstract}


\section{Introduction}
The H$_3^+$ molecular ion is a key molecule for the gas phase
chemistry
in interstellar clouds (Herbst \& Kemplerer 1973; Watson 1973).
Produced by the
fast reaction of H$_2$ with H$_2^+$, it reacts with almost all
neutral atoms and molecules, and thus it is the precursor of a large number
of complex
neutral and ionic molecular species.
Initially suggested to be present in molecular clouds by Martin et al. (1961),
there is a long history behind the search for this crucial molecule. 
Due to its lack of permanent dipole moment,
H$_3^+$ must be observed
in the near--IR through its ro-vibrational transitions (Oka 1980). 
The first searches only provided upper limits to the H$_3^+$ column density  
(Geballe \& Oka 1989; Black et al., 1990). However, H$_3^+$ has been finally
detected
toward GL2136 and the W33A clouds (Geballe \& Oka 1996). 
In addition, H$_3^+$ has also been detected toward the diffuse
interstellar medium (ISM) by McCall et al. (1998), 
and more recently in the line of
sight toward several galactic center clouds (Oka et al. 2005).
Unfortunately, these observations
only provide information on the H$_3^+$ abundance in the
most diffuse and warm external
layers of molecular clouds, and are limited to lines of sight with bright
near-IR background sources. 

A different approach to study H$_3^+$ in denser regions
obscured by dust extinction
is to observe the asymmetrical species H$_2$D$^+$ and
HD$_2^+$, which do have a rotational spectrum.
These species play a crucial role
in the deuteration enhancement of many molecular species in clouds
with kinetic temperatures below $\simeq$30~K (see, e.g., Gerlich et al., 2002 and
references therein).
In particular, the H$_2$D$^+$/H$_3^+$ abundance
ratio is expected to dramatically increase above the elemental
D/H ratio as the gas cools down.
The first calculation of the H$_2$D$^+$ rotational spectrum
was reported by Dalgarno et al. (1973), who suggested
that it could be detectable in dense and cold clouds.
The H$_2$D$^+$ 1$_{1,0}$-1$_{1,1}$ transition has been observed
in the laboratory at 372.42134 GHz (Bogey et al. 1984; Warner et al. 1984). 
This precise measurement opened the quest for the detection of H$_2$D$^+$
in dense molecular clouds. Initial searches
for the o-H$_2$D$^+$ 1$_{1,0}$-1$_{1,1}$ line, and also the p-H$_2$D$^+$
1$_{0,1}$-0$_{0,0}$
(at  1370~GHz), performed with the $\textit{Kuiper Airborne Observatory}$ 
(Phillips et al., 1985; Pagani et al., 1992a; Boreiko \& Betz, 1993),
and with the $\textit{Caltech Submillimeter Observatory}$ (van Dishoeck et al., 1992)
were, however, unsuccessful.
Fortunately, H$_2$D$^+$ has finally been detected
toward the young low-mass protostar
NGC1333-IRAS4A (Stark et al. 1999), and in several
dark clouds (Caselli et al. 2003; Vastel et al., 2004, 2006a,b;
Hogerheijde et al., 2006; Harju et al., 2006), and tentatively
in protoplanetary
disks (Ceccarelli et al., 2004; see also Guilloteau et al., 2006).

Observational evidence for the crucial role of H$_2$D$^+$, HD$_2^+$ 
(Vastel et al., 2004), and even of D$_3^+$,
in the deuteration processes of the ISM have been widely probed
by the detection of double and triple deuterated species
such as D$_2$CO (Turner, 1990; Ceccarelli et al., 1998), NHD$_2$
(Roueff et al., 2000; Loinard et al., 2001; Gerin et al. 2006),
ND$_3$ (van der Tak et al.,
2002; Lis et al., 2002), CHD$_2$OH (Parise et al., 2002), CD$_3$OH (Parise
et al., 2004), D$_2$CS (Marcelino et al., 2005).
\clearpage

Ground-based observations of H$_2$D$^+$ and HD$_2^+$ have always been
performed toward very dense and cold cores as the high Einstein coefficients
of the involved transitions require high volume densities to produce significant
emission. Therefore, it is difficult to estimate the H$_2$D$^+$ abundance
in lower density regions through submillimeter observations.
In this Letter, we report on the first detection of H$_2$D$^+$
in the far-IR spectrum of Sgr B2 observed
with the \textit{Infrared Space Observatory} (ISO).
The observed transition connects the ground state, 1$_{1,1}$, with
the 2$_{1,2}$ level at 113.4 K, and has a wavelength of 126.853 $\mu$m.
This detection opens the possibility
to detect H$_2$D$^+$ in clouds with moderate volume density in which
the 1$_{0,1}$-1$_{1,1}$ transition at 372 GHz will be subthermally
excited and its intensity too weak to be detected with currently available
radio telescopes. It also opens the quest to detect H$_2$D$^+$
toward far-IR luminous galaxies and high redshift objects.

\section{Observations}
The data presented in this paper are based on the final calibration
analysis of the high spectral resolution line survey of Sgr B2 obtained with
the Long Wavelength Fabry-P\'{e}rot Spectrometer (LWS/FP) on board ISO
in its L03 mode
(Polehampton 2002; Polehampton et al., 2007). The data for NH$_2$, which
are necessary to interpret the H$_2$D$^+$ absorption reported in this
paper, were previously presented by Goicoechea et al. (2004).
The spectral resolution
of the LWS/FP spectrometer is $\simeq$ 35~km~s$^{-1}$.
The telescope was pointed toward
$\alpha(2000)$=17$^h$ 47$^m$ 21.75$^s$;
$\delta(2000)$=-28$^o$ 23' 14.1''. The H$_2$D$^+$ and NH$_2$ data are presented
in Figs. 1 and 2.

The procedure to reduce the LWS/FP data have been described in detail by
Polehampton et al. (2007). This provides a significant improvement over
the previous reduction of Goicoechea et al. (2004), who did not apply any
interactive processing to the standard pipeline data products except for
the removal of glitches, averaging of individual scans and removal of
baseline polynomials. The improvements applied in the full spectral survey
reduction include determination of accurate detector dark currents (including
stray light), improved instrumental response calibration (including a
correction to account for adjacent FP orders) and interactive shifting of
individual miniscans (to correct for uncertainty in the angle of the LWS
grating). The signal-to-noise ratio was also enhanced by including every
available observation in the survey (in this case 2 independent
observations).
The broad absorption feature shown in Figure 1 was previously assigned
only to p-NH2 (Goicoechea et al., 2004; transition 2$_{2,1}$-1$_{1,0}$
J=5/2-3/2).
However, the final reduction of the LWS/FP survey toward Sgr B2 shows that
the broadening is real. The feature appears in 2 independent observations :
ISO TDT number 50601112, observed on 1997 April 5 with LWS detector LW2;
and ISO TDT number 50800515, observed on 1997 April 7 with LWS detector LW3.
The fact that the feature appears on 2 separate detectors rules out any
spurious instrumental broadening of the p-NH2 line.

Taking into account the latest data reduction,
the broad feature detected at $\sim$126~$\mu$m cannot be due to p-NH$_2$ alone.
In fact, NH$_2$ far-IR lines are narrow and appear only at the velocity
of Sgr B2 (see, van Dishoeck et al., 1993, for
the detection of NH$_2$ in the submm).
\begin{figure}[t]
\centering
\includegraphics[angle=0,width=7.cm]{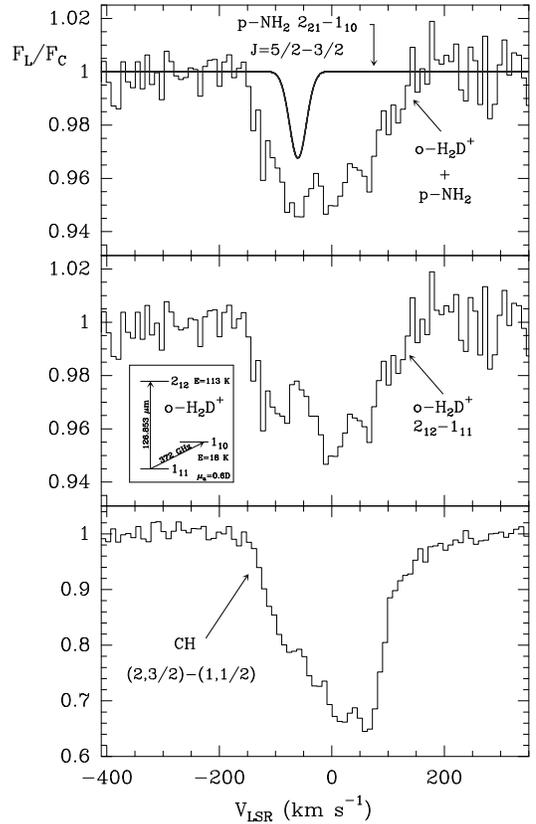}
\caption{Observed ISO spectrum at 126.853 $\mu$m toward Sgr B2. The upper
panel shows the data and the expected contribution of p-NH$_2$ (see Figure 2).
The middle panel shows the o-H$_2$D$^+$ line profile after subtracting
the contribution from p-NH$_2$. Finally, the lower panel shows the
absorption of CH for comparison purposes.} 
\label{fig:figure1}
\end{figure}

Only a few lines remain unidentified in the ISO far-IR line survey of SgrB2.
The feature discussed in this paper must correspond to a line
arising from a low energy level or to a blend of lines. However,
molecules heavier than H$_2$D$^+$ could have other
lines in the far-IR domain.
We have considered the possibility of b-type transitions
from slightly asymmetric rotors like HNCO and HOCO$^+$. Both have one
transition at similar wavelengths. However, the line of HNCO involves two
high energy levels and can be easily ruled out. HOCO$^+$ has several other
lines
in the 120-127 $\mu$m range involving low energy levels. None of which are
detected in the far-IR line survey of Sgr B2 (Polehampton et al., 2007).
Hence, we are confident that the broad feature detected at
126.853 $\mu$m corresponds to H$_2$D$^+$.

The intensity scale in figures 1 and 2 was obtained by dividing the observed
flux by a fitted continuum as described by Polehampton et al. (2007). Hence,
it is directly related with the line opacity convolved with the
instrumental spectral profile of the LWS/FP spectrometer.
The velocity scale in Figure 1
has been obtained assuming $\lambda_{rest}$=126.853~$\mu$m (2363.325 GHz),
i.e., the wavelength of the o-H$_2$D$^+$ 2$_{12}$-1$_{11}$ transition
(Amano \& Hirao, 2005).
The expected error on the frequency of this transition is $\simeq$ 5 MHz,
i.e., a velocity uncertainty $<$ 2 km s$^{-1}$ (3$\sigma$),
which is much lower than the spectral resolution of the LWS/FP spectrometer.
The p-NH$_2$ 2$_{2,1}$-1$_{1,0}$ J=5/2-3/2 line is separated by
$\simeq$-123 km s$^{-1}$ with respect to
the 2$_{12}$-1$_{11}$ line of o-H$_2$D$^+$.
\clearpage

\section{Results and Discussion}
The upper panel of Figure 2 shows the 
three spin--rotational components of the o-NH$_2$ 2$_{2,0}$-1$_{1,1}$
transition at $\sim$117~$\mu$m (Goicoechea et al. 2004).
In order to compute the expected contribution of the p-NH$_2$ 
2$_{2,1}$-1$_{1,0}$ triplet at $\simeq$126 $\mu$m we have modeled the
NH$_2$ absorption
assuming a rotational temperature of 20~K, an intrinsic linewidth of
10 km s$^{-1}$,
a total NH$_2$ column density  of 1.2$\times$10$^{15}$~cm$^{-2}$
(Goicoechea et al. 2004),
and an ortho/para
ratio of 3. 
NH$_2$ wavelengths and line strengths are from the
Cologne Database for Molecular Spectroscopy (M\"uller et al. 2001).
The NH$_2$ hyperfine structure is unresolved by the LWS/FP.
The synthetic spectrum, convolved with the LWS/FP spectral resolution,
is shown in Figure 2. The expected
contribution from p-NH$_2$ to the absorption feature is shown as a continuous
line in the upper panel of Figure 1. The middle panel of this Figure shows
the same data with the expected contribution from p-NH$_2$ removed.
The remaining absorption feature is broad and has several absorption velocity peaks,
which are similar
to those found in unsaturated ground--state transitions of light species
such as the (2,3/2)-(1,1/2) line of CH (see bottom panel of Figure 1).
Therefore, in the following, we assign this unidentified broad feature
to the o-H$_2$D$^+$ 2$_{1,2}$-1$_{1,1}$ line.

H$_2$D$^+$ is an asymmetric molecule with $\mu_a$ $\simeq$0.6 D
(Dalgarno et al., 1973). It has two different species, ortho
(Ka=odd), and para (Ka=even),
with a spin weight ratio of 3:1.
The three lowest energy levels, 1$_{1,1}$, 1$_{1,0}$ and 2$_{1,2}$, which are
at 0, 17.9 and 113.4 K
(see insert in the middle panel of Figure 1),
are enough to compute the partition function
(F$_{par}$) at low temperatures.
Hence,
F$_{par}(T_r)$ $\simeq$ 3(1+e$^{-17.8/T_r}$) is a good
approximation for T$_r$ $<$ 30 K, where T$_r$ is the rotational
temperature.
For example, F$_{par}$(3 K)=
3.004 (all molecules in the ground state), while F$_{par}$(30 K)=4.8.
Although the 1$_{1,0}$ level
starts to be populated in this case, the 2$_{1,2}$ one is
still unpopulated. Consequently, if
there is a strong continuum
source behind or inside the source, we could expect to see
the 2$_{1,2}$-1$_{1,1}$ transition of o-H$_2$D$^+$ in absorption.
The same behavior is observed in CH toward
Sgr~B2, where only the ground-state line is observed
with the line profile shown in the bottom panel of Figure~1 (Goicoechea et
al. 2004; Polehampton et al. 2007). 

The Einstein coefficient of the o-H$_2$D$^+$ 2$_{1,2}$-1$_{1,1}$
transition  is large,
1.7$\times$10$^{-2}$ s$^{-1}$, which implies that
the rotational temperature of the line will be extremely low
in most relevant cases
(or close to the dust temperature if continuum emission is
optically thick),
even when the 1$_{1,0}$-1$_{1,1}$ transition is already
collisionally excited (its Einstein coefficient is two orders of
magnitude lower).
Therefore, in order to derive a column density
from absorption measurements of the 2$_{1,2}$-1$_{1,1}$ line,
we can assume that all molecules are in the 1$_{1,1}$ ground
level, and that the transition has a low rotational temperature
in absence of infrared pumping.
The opacity of
the 2$_{1,2}$-1$_{1,1}$ line can be written as $\tau(T_r)\simeq$
0.067(N/10$^{13}$)(10/$\Delta$v)/F$_{par}(T_r)$,
where $\Delta$v is the linewidth at half intensity in km s$^{-1}$, and
N is the column density of o-H$_2$D$^+$ in cm$^{-2}$.
If N(o-H$_2$D$^+$)=10$^{13}$ cm$^{-2}$ and $\Delta$v=10 km s$^{-1}$,
$\tau$(2$_{1,2}$-1$_{1,1}$) would be 0.024 for T$_r$=3~K and 0.014 for T$_r$=30~K
respectively.
For comparison purposes, the opacity of the 1$_{1,0}$-1$_{1,1}$ transition
at 372 GHz would be 0.024 and 0.006 for T$_r$=3 and 30~K respectively.

The interpretation of the absorption produced by o-H$_2$D$^+$ in Sgr~B2
requires knowledge of the rotational temperature of the transition. The
cloud has a large dust opacity in the far-IR,
$\tau_d$(100$\mu$m)=3.5 (Cernicharo et al. 1997, 2006; Goicoechea et al. 2004).
Hence, infrared photons cloud play an important role in the
pumping of the molecular levels of o-H$_2$D$^+$.
In addition, the fraction of the cloud depth that is "seen" at this 
wavelength corresponds only
to its external layers, where $\tau_d$(127 $\mu$m)$\simeq$1, i.e.,
N(H$_2$)$\simeq$3$\times$10$^{23}$ cm$^{-2}$, rather than the
$\geq$10$^{24}$ cm$^{-2}$
corresponding to the total column density along the line of sight
($\tau_d$(127 $\mu$m)$\simeq$3).
In order to estimate the radiative and collisional
effects on the excitation of the
2$_{1,2}$-1$_{1,1}$
transition, we have assumed de-excitation collisional rates 
$\sigma$(1$_{1,0}$-1$_{1,1}$)=$\sigma$(2$_{1,2}$-1$_{1,0}$)=
10$^{-10}$ cm$^3$ s$^{-1}$,
and $\sigma$(2$_{1,2}$-1$_{1,1}$)=5$\times$10$^{-11}$ cm$^3$ s$^{-1}$.
For a gas at 30 K purely excited by collisions with N(o-H$_2$D$^+$)=
10$^{13}$ cm$^{-2}$, and $\Delta$v=10 km s$^{-1}$,
the rotational temperature varies
from 9 to 12 K for H$_2$ densities of 10$^5$ and 10$^6$ cm$^{-3}$
respectively. Reducing the collisional rates by a factor 10
produces rotational temperatures of 6 and 8 K for the same volume densities.

\begin{figure}[t]
\centering
\includegraphics[angle=-90,width=8.5cm]{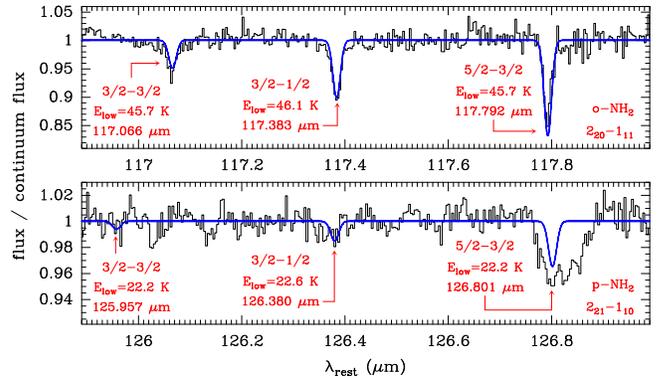}
\caption{Rest ISO/LWS/FP  spectrum of Sgr~B2 around 117 and 126 $\mu$m,
corrected by the averaged velocity of far--IR lines toward the source
(62.7 km s$^{-1}$).
Lines due to o-NH$_2$ (upper panel) and p-NH$_2$ plus o-H$_2$D$^+$
(lower panel) are shown. The expected NH$_2$ absorption  is
shown by continuous lines in both panels (see text). An NH$_2$ ortho/para ratio of 3 has
been assumed. Note that the NH$_2$ lines
show absorption only at the velocity of Sgr~B2.}
\label{fig:figure2}
\end{figure}

The role of radiative pumping by far--IR
dust photons can be estimated assuming that
the gas we observe in absorption, $\tau_d$(127$\mu$m)=1, surrounds
a region with $\tau_d$(127$\mu$m)=2.0 and T$_d$=30 K.
If the radius of the absorbing shell is twice that
of the region emitting the bulk of the infrared emission,
R$_{abs}$/R$_{cont}$=2, we obtain an excitation temperature of 17 K, with a
little dependency on the assumed collisional rates.
For R$_{abs}$/R$_{cont}$=3 and 1.5, the rotational
temperature is 15 and 20 K respectively.
In the following we assume T$_r$=10-20 K, and we use
these values  to derive upper and lower limits
to the opacity of the H$_2$D$^+$ 2$_{1,2}$-1$_{1,1}$ line.
Assuming a linewidth of 10~km~s$^{-1}$ we obtain
$\tau$(2$_{1,2}$-1$_{1,1}$)=0.18 and 0.36 for T$_r$=10 and 20 K
respectively. The corresponding column densities are $\simeq$
9$\times$10$^{13}$ and 2.3$\times$10$^{14}$ cm$^{-2}$, and the
abundance of o-H$_2$D$^+$, for a total gas column
density of 3$\times$10$^{23}$ cm$^{-2}$, is
$\simeq$ 3$\times$10$^{-10}$ for T$_r$=10 K
and 9$\times$10$^{-10}$ for T$_r$=20 K. The latter 
can be ruled out
from the upper limit obtained by Pagani et al (1992a)
for the emission of the 1$_{1,0}$-1$_{1,1}$ line at 372 GHz,
N(o-H$_2$D$^+$)$<$1.1$\times$10$^{14}$~cm$^{-2}$. Hence, a low rotational
temperature, T$_r\simeq$10 K, and an abundance of 3$\times$10$^{-10}$ are
the best fit to the o-H$_2$D$^+$ absorption at the velocity of
Sgr~B2.

For the other absorbing clouds which are not directly associated to
Sgr~B2, it is reasonable to assume radiative thermalization
with the cosmic background,
T$_r$ $\simeq$3~K, i.e., no significant
collisional excitation or infrared pumping. Assuming
an intrinsic line width of 10 km s$^{-1}$, the observed absorption
features at -100, 10 and 120 km s$^{-1}$ have 
opacities of of 0.1, 0.15 and 0.06 and
column densities of $\simeq$ 4.5, 6.7, and 2.7 10$^{13}$ cm$^{-2}$ respectively.
These clouds have low volume densities and total gas column densities
of a few 10$^{22}$-10$^{23}$ cm$^{-2}$. The associated o-H$_2$D$^+$
abundances are of a few 10$^{-10}$--10$^{-9}$.

The derived abundance of o-H$_2$D$^+$ in Sgr~B2 corresponds to the
predictions of chemical models for T$_K$ $<$20 K (see Pagani
et al., 1992b; Walmsley et al., 2004; Flower et al., 2006).
These models predict that the
para species could be more
abundant than the ortho one for that temperature
and the expected density of the gas (a few 10$^4$-10$^5$ cm$^{-3}$).
Hence, the fundamental
line of p-H$_2$D$^+$ could produce significant absorption at 218.8 $\mu$m
in the direction of Sgr~B2.
The abundances observed toward the other clouds in the line of sight
of Sgr~B2 indicate lower
kinetic temperatures, T$_K\simeq$10 K. In these clouds the predicted
ortho/para abundance ratio favors the ortho species.
Due the high dependency of the ortho/para ratio with the gas temperature,
simultaneous observation of both the 126.853 $\mu$m (o-H$_2$D$^+$) and the
218.8 $\mu$m (p-H$_2$D$^+$) lines
will provide a direct measurement of the
efficiency of the reactions of H$_2$D$^+$ with H$_2$ and of the
conversion of ortho--to--para H$_2$D$^+$ and vice-versa, i.e., to
check chemical models.

Due to the poor spectral and angular resolution of ISO, and due to the limited
sensitivity of the LWS/FP instrument, the o--H$_2$D$^+$ line at 126.853 $\mu$m
could only be expected in sources with a large column density of cold gas
with broad linewidths
in the line of
sight. As an example, a broad absorption feature
has been also reported in the bright IR galaxy Arp220
at 127 $\mu$m by Gonz\'alez-Alfonso et al. (2004) using the LWS grating
spectrometer (spectral resolution more than 20 times lower than the LWS/FP).
These authors assigned
it to NH$_3$. While several lines of NH and NH$_3$ have been
observed in this source, no lines have been assigned to NH$_2$ so far. Hence,
the broad absorption feature assigned to NH$_3$ by Gonz\'alez-Alfonso et al.,
could also contain some contribution from o-H$_2$D$^+$. 
The detection of H$_2$D$^+$ presented in this work, shows that far--IR 
ortho-- and para--H$_2$D$^+$ lines  can be used to trace the deuteration 
fractionation of H$_3^+$ in galaxies (see also Ceccarelli and Dominik,
2006).
Therefore, far--IR observations with the PACS instrument
on board Herschel satellite (or in the future with SOFIA), will be an
important tool
to understand  the cold ISM chemistry in distant galaxies
where the D/H ratio will be different and the gas temperature affected
by the local cosmic background temperature.

Finally, the opacity of the 2$_{1,2}$-1$_{1,1}$ transition for a dark
cloud with $\Delta$v=0.5 km s$^{-1}$ (see Vastel et al., 2004),
N(o-H$_2$D$^+$)=10$^{12}$ cm$^{-2}$, and T$_r$=
3 K is $\simeq$0.05. While the expected emission in the 1$_{1,0}$-1$_{1,1}$
transition at 372 GHz will be negligible, the expected absorption at 126.853
$\mu$m will be 5\%, i.e., easy to detect with the
expected sensitivities of the high spectral
resolution instruments planned for SOFIA.
Dark clouds (see, e.g., Nisini et al., 1999), prestellar cores,
and protoplanetary disks (see Wyatt et al., 2003)
have enough continuum emission
in the far-IR, hence, the 2$_{1,2}$-1$_{1,1}$ transition can be used as a key
observing tool to trace column densities of o-H$_2$D$^+$
larger than a few 10$^{11}$ cm$^{-2}$, and to understand the
deuterium fractionation at large spatial scales
in cold dark clouds.
Observations of the fundamental lines of
p-HD$_2^+$ at 691.6605 GHz and o-HD$_2^+$ 203.03 $\mu$m will provide
additional constraints on these deuteration processes in molecular clouds
and protoplanetary disks.

\acknowledgments
We would like to thank Cecilia Ceccarelli, Serena Viti, and our
referee for useful comments and suggestions.
We also thank the spanish MEC for funding support
through grant ESP2004-665, AYA2003-2785, and ``Comunidad de Madrid''
Government under
PRICIT project S-0505/ESP-0237 (ASTROCAM). This study is
supported in part by the European Community's human potential
Programme under contract MCRTN-CT-2004-51230, Molecular Universe.
JRG was supported by an individual Marie Curie fellowship under contract
MEIF-CT-2005-5153340.

\end{document}